\pdfoutput=1
\documentclass[traditabstract]{aa} 
\usepackage{graphicx}
\usepackage{txfonts}
\begin{document}

\title{Potential-density pairs \\ and vertical tilt of the stellar velocity ellipsoid}
\titlerunning{Velocity ellipsoid tilt}
\author{O. Bienaym\'e\inst{}}
\date{\today}                                           

   \authorrunning{O. Bienaym\'e}
   \offprints{bienayme@astro.u-strasbg.fr}
%
   \institute{
Universit\'e de Strasbourg, CNRS, Observatoire Astronomique, France
}
   \date{09/03/2009}
   \abstract{ We  define new  potential-density pairs and  examine the
impact of the potential  flattening on the vertical velocity ellipsoid
tilt,  $\delta$.  By  means of  numerical integrations  and analytical
calculations,  we   estimate  $\delta$   in  a  variety   of  galactic
potentials.  We  show that at  1 kpc above  the Galactic plane  at the
solar radius, $\delta$  can differ by 5 degrees,  depending on whether
the dark matter  halo is flat or spherical.   This result excludes the
possibility of an extremely flattened Galactic dark halo.

         \keywords{gravitation -- 
Galaxy: kinematics and dynamics  --}

}
   \maketitle

\section{Introduction}

The  velocity  ellipsoid shape  can  be  determined  from the  stellar
kinematics in our Galaxy, see  for instance Chereul et al. (\cite{che98}) from
Hipparcos  data, Soubiran  et al.  (\cite{sou08}) from  distant red  giants or
Siebert et al.  (\cite{sie08}) from Rave data.  It is  known that the vertical
tilt   of  the  velocity   ellipsoid  is   related  to   the  galactic
gravitational  potential  and  that,  in  the  peculiar  case  of  the
St\"ackel potentials, it only  depends on the gravitational potential.
Probably because  of the  lack of data,  very few works  have examined
this question since  the pioneering work of Ollongren  (\cite{oll62}).  We can
note  the  work  by  Kuijken  and  Gilmore  (\cite{kui89})  that  proposes  an
approximate  determination of  the inclination  at the  solar position
above the galactic plane and that points out the necessity to consider
this  inclination to  determine the  $K_z$ force  out of  the galactic
plane.  There are also the  analytical studies by Cuddeford and Amendt
(\cite{cud91}) of the Jeans equations to the fourth order and their analytical
approximation  of  the  vertical  tilt.  Many  other  studies  concern
St\"ackel  potentials for  which the  ellipsoid orientation  is easily
obtained.

 Thus for  realistic galactic  potentials and from  orbit integration,
both  Kuijken   \&  Gilmore  (\cite{kui89})  and   Binney  \&  Spergel
(\cite{bin83}) show that at 1\,kpc out of the galactic plane above the
Sun, the velocity ellipsoid points towards  a point located at 5 to 10
kpc  behind  the   galactic  center  (see  also  Kent   \&  de  Zeeuw,
\cite{ken91}, and Shapiro et al., \cite{sha03}).

  In  Section 2,  we define  a new  potential-density pair  to measure
(Section 3)  the vertical tilt  of the velocity  ellipsoid numerically
within a  spheroidal potential by varying its  flattening.  In Section
4,  we  generalize the  potential-density  pair  to a  three-parameter
family with  properties similar to the Miyamoto  and Nagai potentials,
but with flat circular velocity  curves at large radii.  In Section 5,
we define  a simple  expression to estimate  the tilt and  examine its
accuracy.

\section{A  new potential-density pair}

 We make the  definition of the Mestel disk  more general by utilizing
the scale-free axisymmetric potential
\begin{equation}
 \phi(R,z)= v_c^2 \ln \left( \sqrt{R^2+z^2} + \sqrt{q^2 R^2+z^2}  \right).
\label{eq:q}
\end{equation}
 It generates, within  the plane of  symmetry $z$=0, a  constant circular
velocity  curve  of  amplitude   $v_c$.  
The resulting density  is always positive:

\begin{equation}
 \rho(R,z)= \frac{v_c^2}{4\pi G}
 \frac{ q^2 \left( R^2+z^2 \right)^{1/2} }{ \left(q^2 R^2+z^2 \right)^{3/2}}.
\label{eq:dq}
\end{equation}

\begin{figure}[!htbp]
\resizebox{\hsize}{!}{
\includegraphics[angle=0] {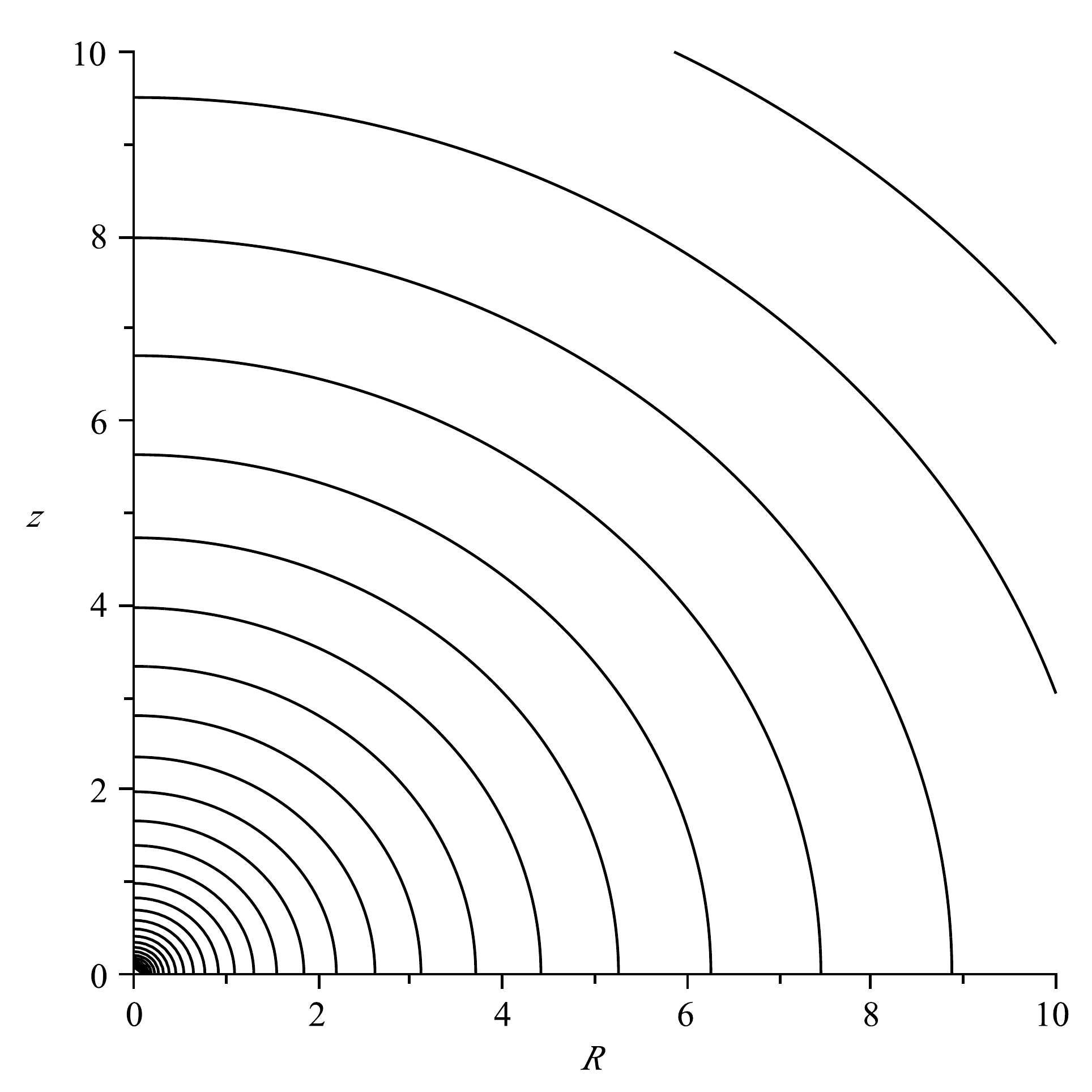}
\includegraphics[angle=0] {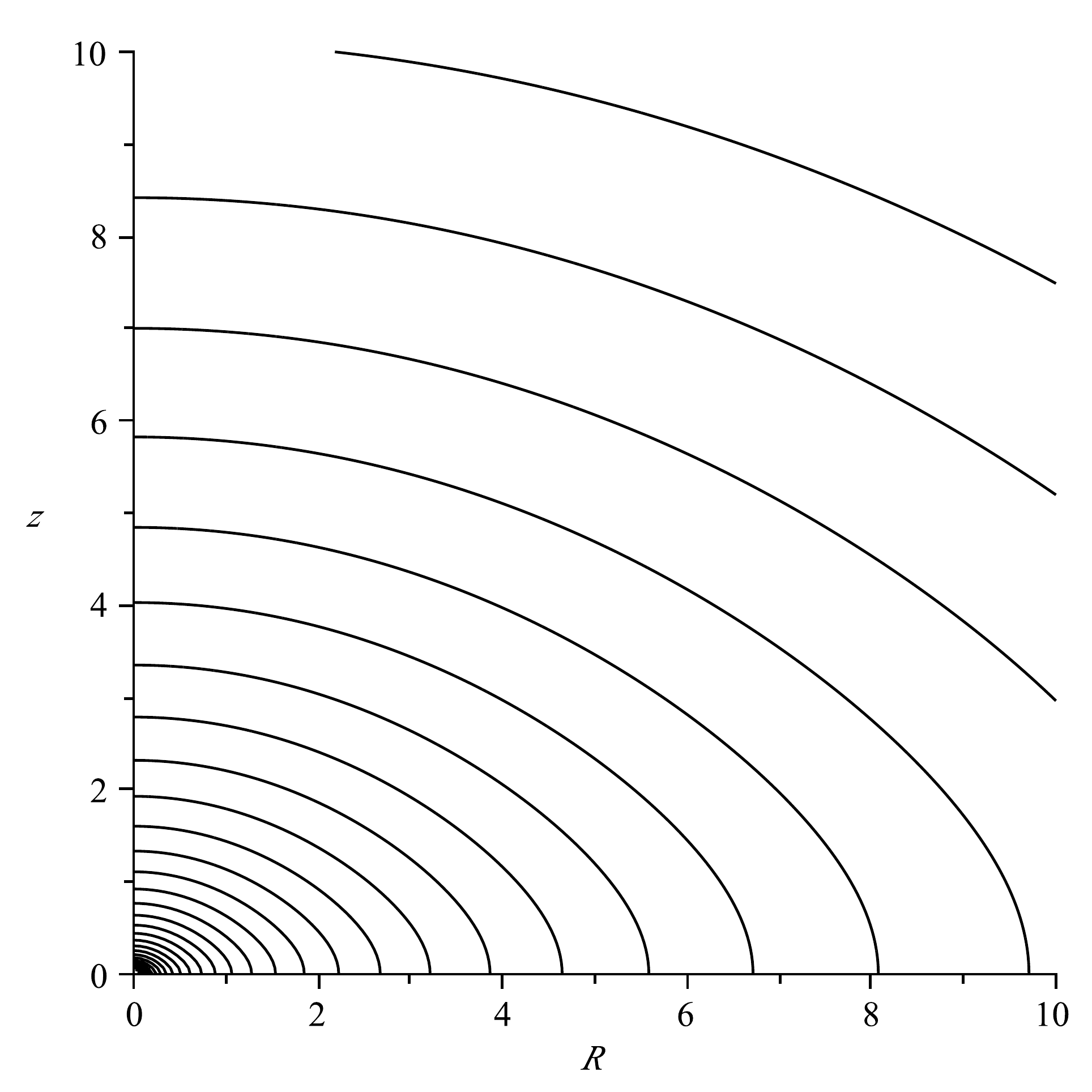}
}
\caption{  Contours of isopotentials with $q$=0.8 and 0.2.}
\label{f:pot1}
\end{figure}

\begin{figure}[!htbp]
\resizebox{\hsize}{!}{
\includegraphics[angle=0] {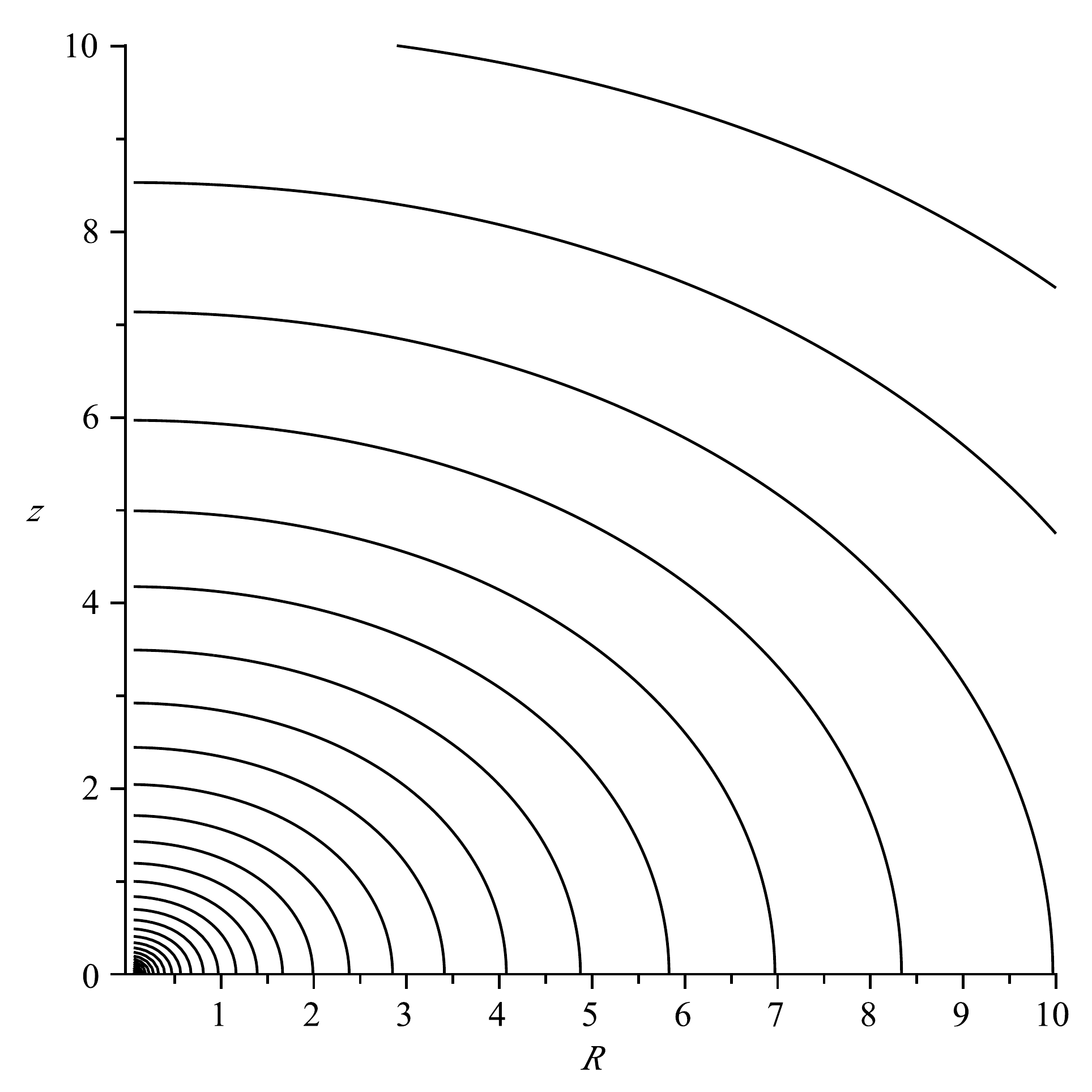}
\includegraphics[angle=0] {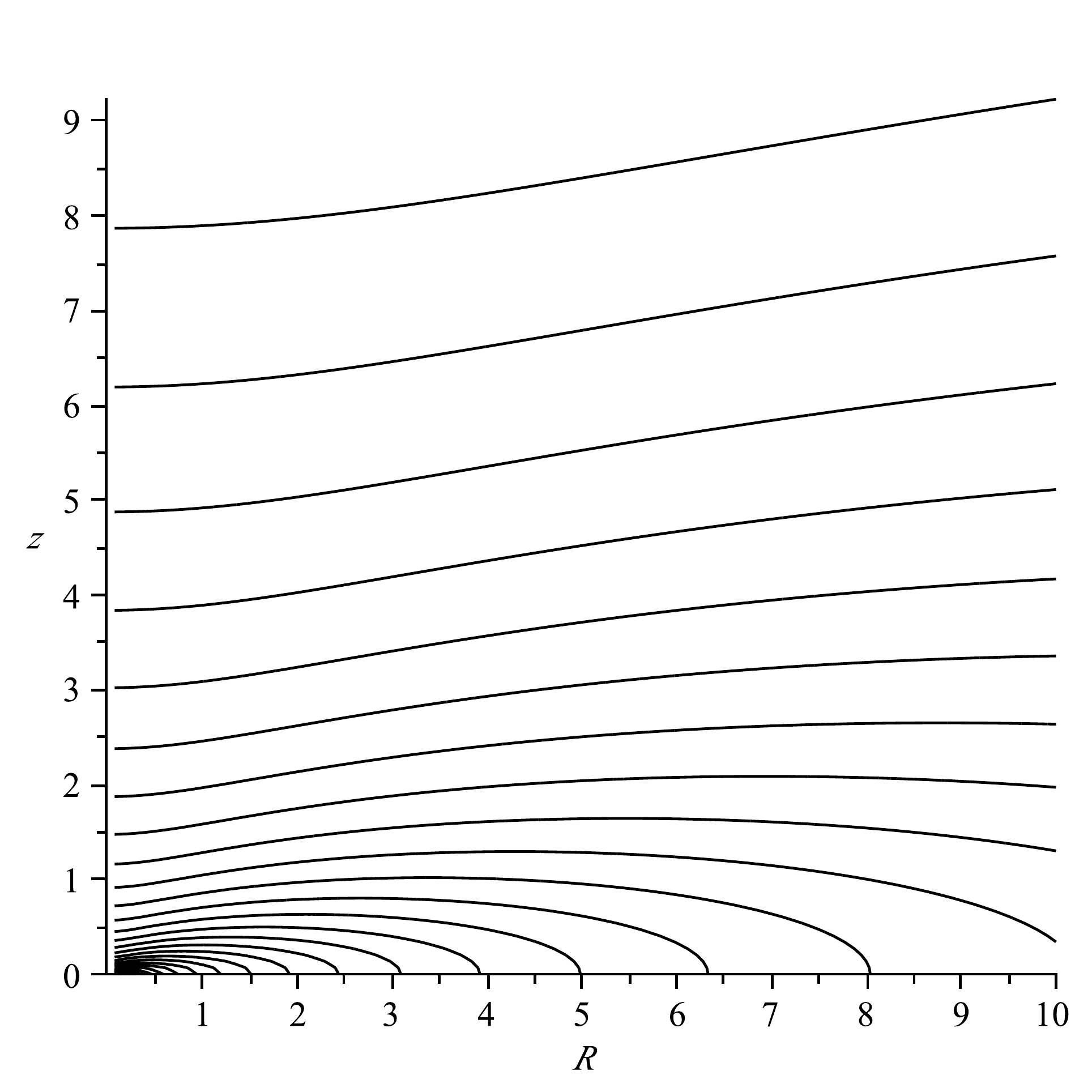}
}
\caption{  Contours of isodensities with $q$=0.8 and 0.2.}
\label{f:dens1}
\end{figure}

Figures    \ref{f:pot1}-\ref{f:dens1}   show   the    isodensity   and
isopotential  contours for  some values  of  $q$.  The  axis ratio  of
isodensity  contours is  $q^{3/2}$  and for  the  isopotentials it  is
$(1+q)/2$,   never  less  than   1/2.   It   can  be   shown  (Toomre,
\cite{too82})  that any  scale-free potential  at the  spherical limit
(obtained here with $q$=1) is the potential of the singular isothermal
sphere, while  the flat  disk limit ($q$=0)  is the Mestel  flat disk.
With  $q$$ >$1,  the Eq.~\ref{eq:q}  potential is  prolate and  at the
limit  $q$=$ \infty$ the  potential and  density distributions  do not
depend on the $z$ coordinate (cylindrical distribution).

\section{Velocity ellipsoid tilt}
\begin{figure}[!htbp]
\resizebox{\hsize}{!}{
\includegraphics[angle=-90] {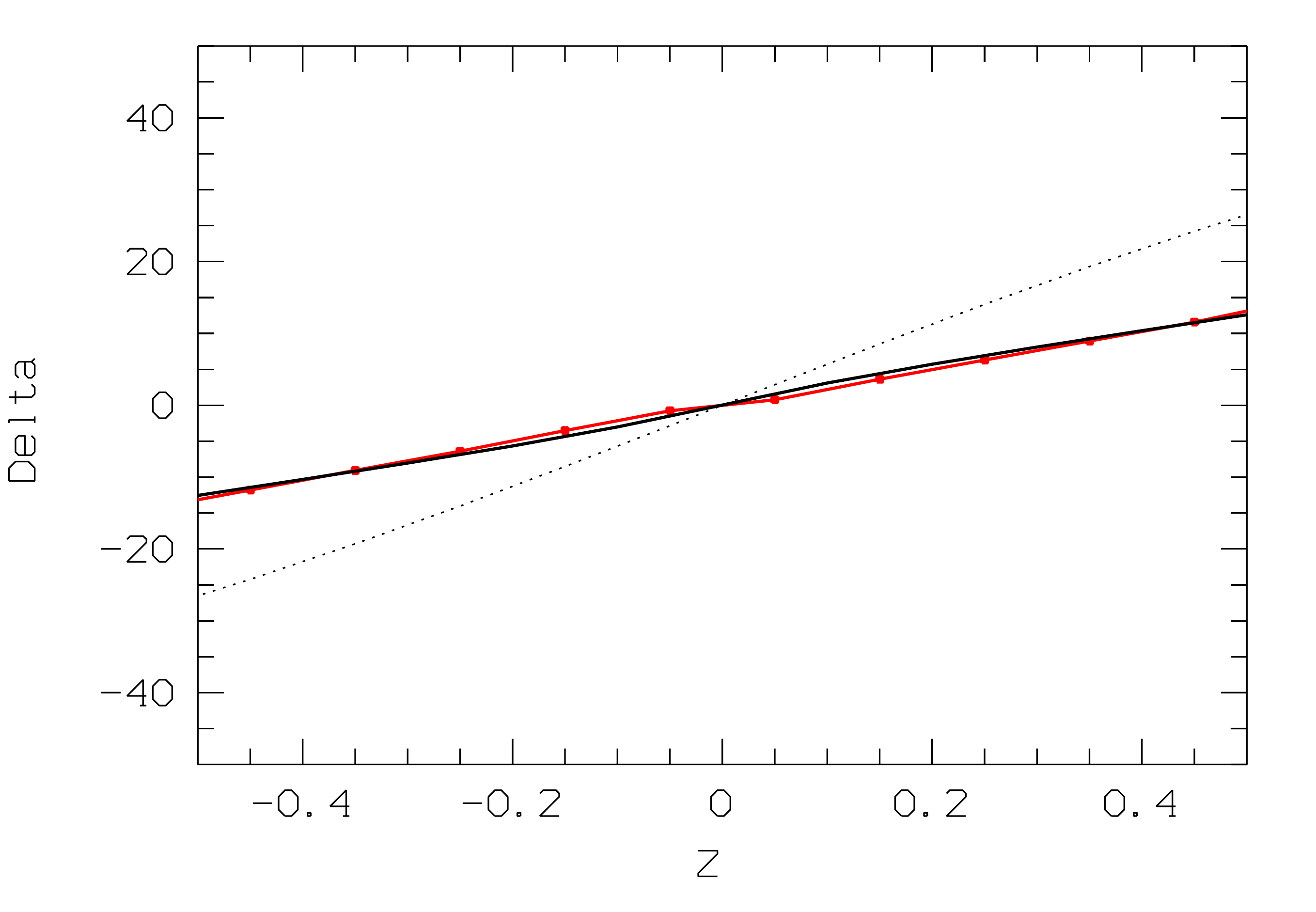}
\includegraphics[angle=-90] {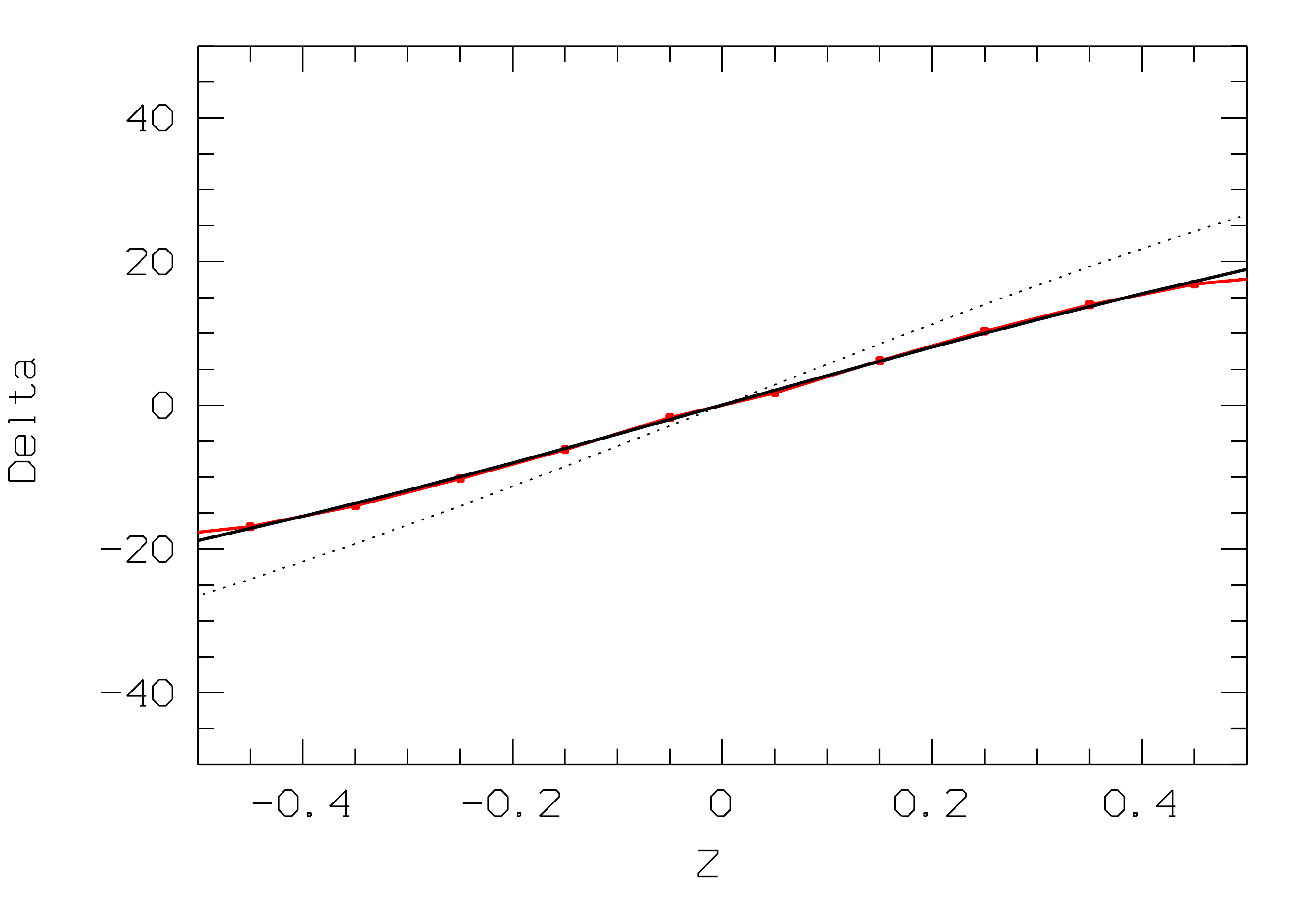}
}
\caption{Vertical tilt of the  velocity  ellipsoid   $\delta$  in  degrees
versus  $z$  at $R=1$  for  models  with  $q=0.3$ (left)  and  $q=0.7$
(right).    Black line:   analytical estimate
(see Section  5),    red  line:    result  of  numerical  orbit
integration,    dotted line:   tilt within a   spherical potential.}
\label{f:tilt1}
\end{figure}

Since the  Eq.\,\ref{eq:q} potential is  scale-free we simply  need to
determine the velocity ellipsoid  orientation versus $z$ at some given
radius $R$.   We build the  velocity ellipsoid numerically  from orbit
integrations  (2.\,10$^7$  orbits  with $\sigma_z/\sigma_R\sim$1/2  at
$z$=0).   The velocity  ellipsoid orientation  for  these distribution
functions is shown (Fig.  \ref{f:tilt1}), for $R$=1, versus $z$ within
the  meridional plane  for potentials  with flattenings  $q$=0.3, 0.7.
Other approaches  could be  tried: see for  instance, the  analysis by
Evans et  al. (\cite{eva97}) from  the Jeans equations  for scale-free
potentials with a flat rotation curve by applying an approximate third
integral   (see  also   for   scale-free  potentials:   Qian  et   al.
\cite{qia95},  Evans  \&  de  Zeeuw \cite{eva92},  Qian  \cite{qia92},
Hunter et  al.  \cite{hun84}, de  Zeeuw et al. \cite{dez96},  Evans et
al. \cite{eva97}).

\begin{figure}[!htbp]
\centering
\resizebox{0.8\hsize}{!}{
\includegraphics[angle=-90] {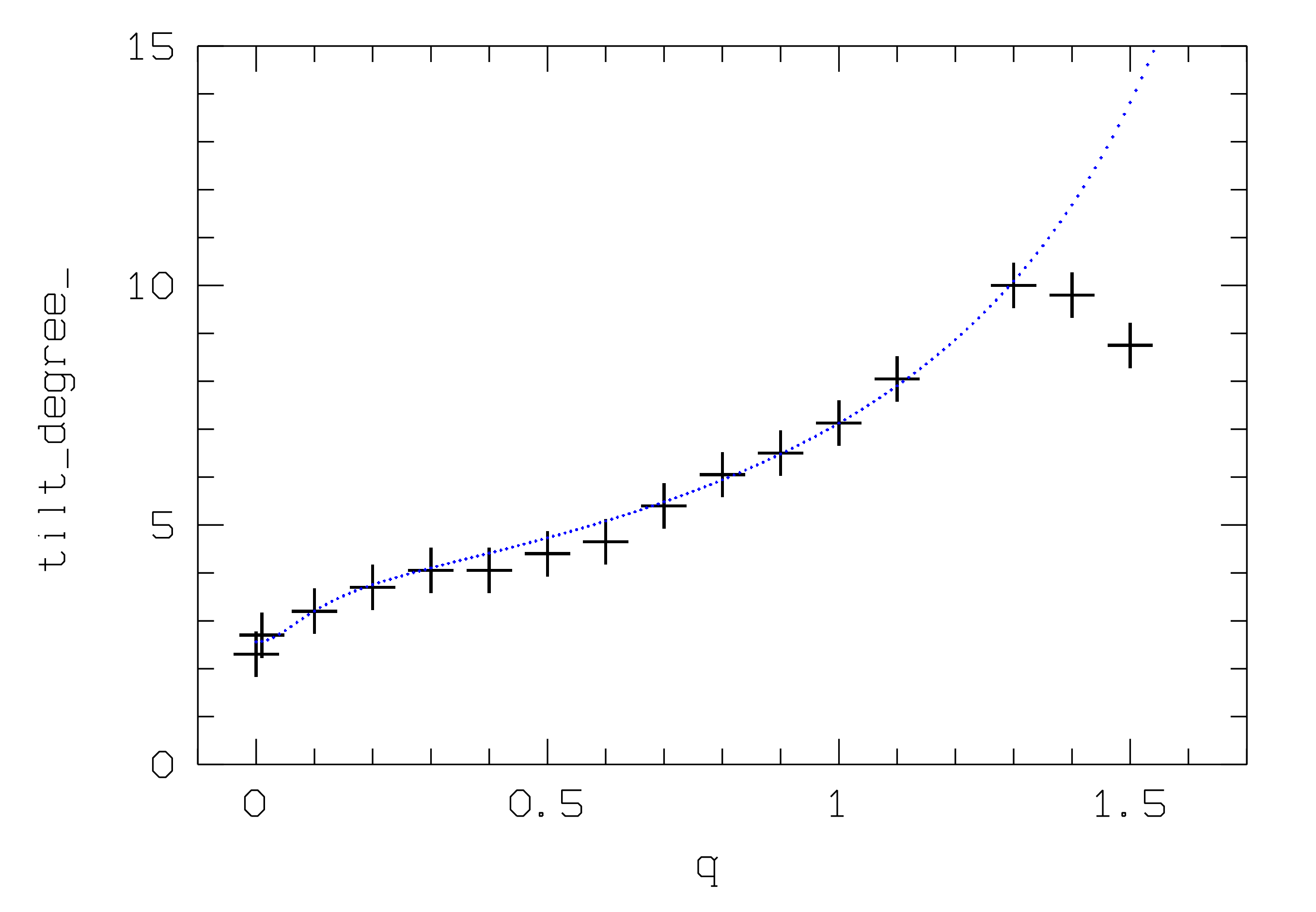}
}
\caption{Velocity   ellipsoid  inclination   in  degrees   versus  
potential flattening $q$  at $R$=8 and $z$=1.   Dotted line: the
expected   inclination  from   Section  5,   crosses:  numerical
determinations from orbit integration.}
\label{f:ivzq}
\end{figure}

Figure \ref{f:ivzq} shows the variation  in the tilt angle $\delta$ at
$R$=8  and  $z$=1 above  the  mid-plane  versus  the coefficient  $q$.
$\delta$ varies by 5 degrees from 2.2 to 7.1 degrees {\ depending on }
whether the  density distribution is  flat or spherical.  This  may be
compared with the recent determination by Siebert et al (\cite{sie08})
of  the inclination  of  the velocity  ellipsoid  at 1  kpc below  the
Galactic plane.   From RAVE data (Zwitter et  al.  \cite{zwi08}), they
find  an inclination $\delta=7.3\pm1.8$  degrees towards  the Galactic
plane.  They show it implies a moderate flattening of the dark halo by
exploring  the   velocity  ellipsoid  tilt   within  various  galactic
potentials (Dehnen  \& Binney \cite{deh98}).  Within  the potential of
the Besan\c  con galactic model  (Robin et al  \cite{rob03}) (probably
the  best  current  Galactic  baryonic  mass model)  that  includes  a
spherical  dark halo, we  also determine  the velocity  ellispoid tilt
from  numerical  orbit integration  and  obtain $\delta\sim6$  degrees
(Bienaym\' e  et al. \cite{bie09}).  To summarize,  within a spherical
Galactic potential,  the tilt at  1 kpc above  the plane at  the solar
radius  is about  7 degrees,  and an  extreme flattening  of  the dark
matter component is excluded, since the  tilt would be close to 2 or 3
degrees (Fig. \ref{f:ivzq}).

\section{Other potential-density pairs}

\begin{figure*}[!htbp]
\centering
\resizebox{17cm}{!}{
\includegraphics[angle=0] {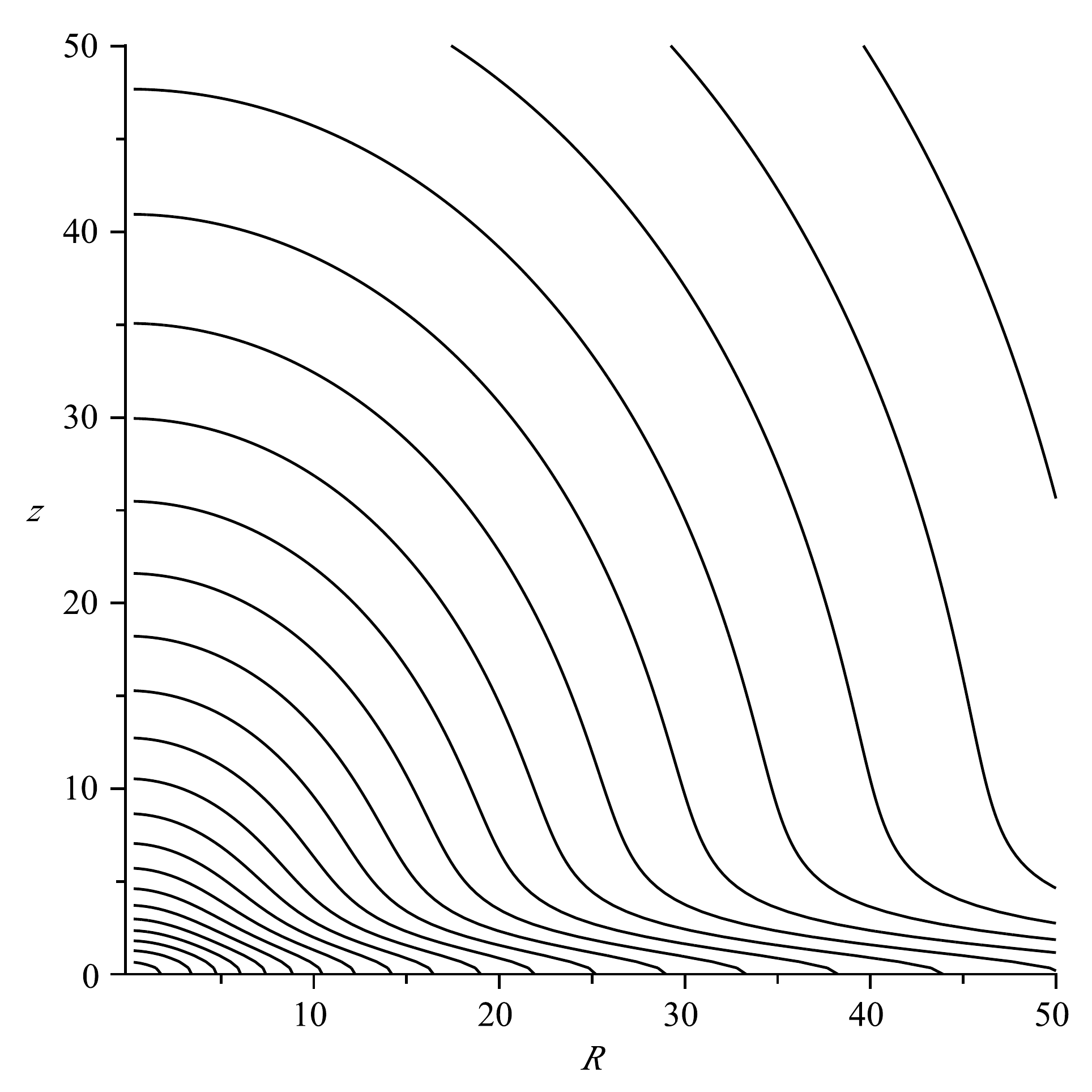}
\includegraphics[angle=0] {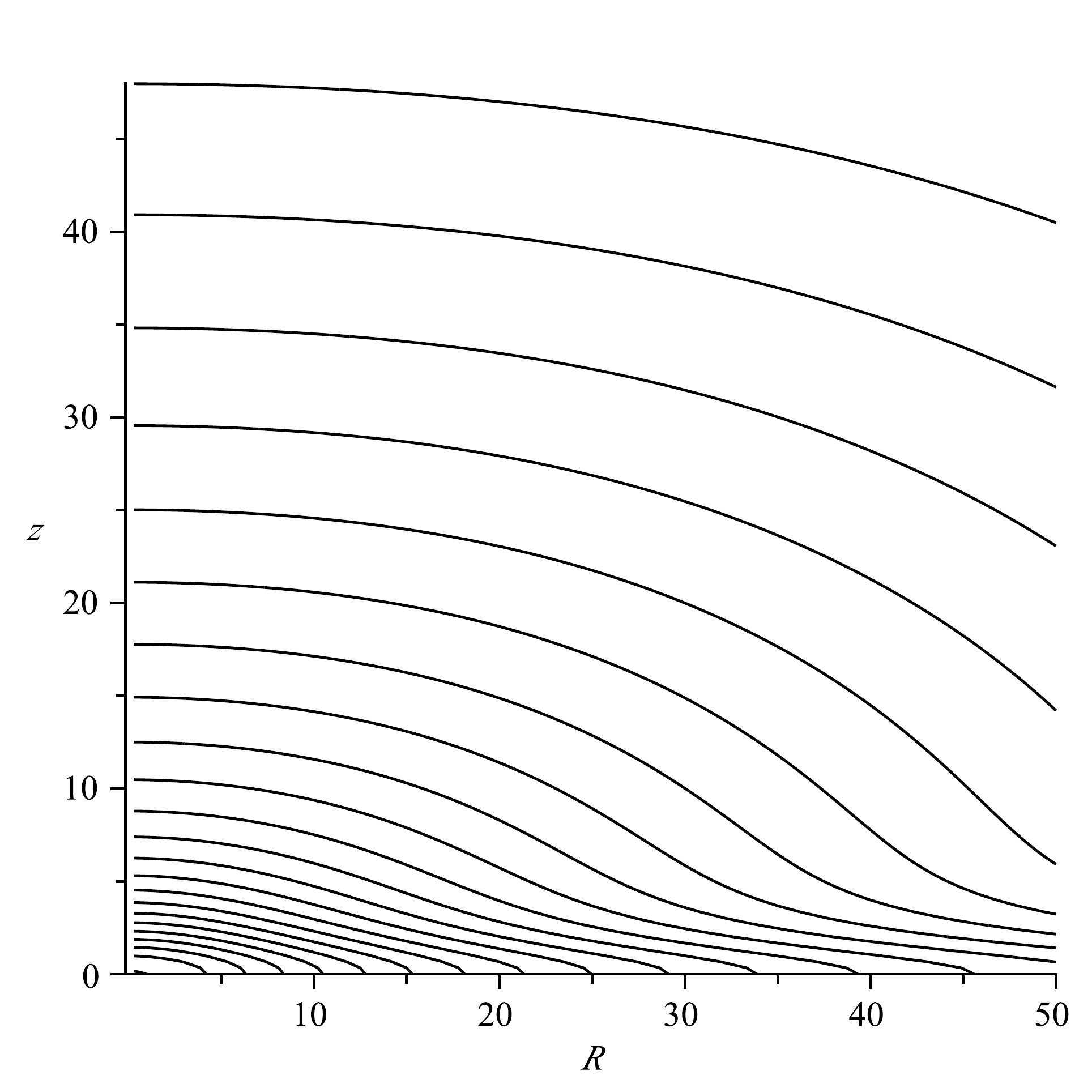}
\includegraphics[angle=0] {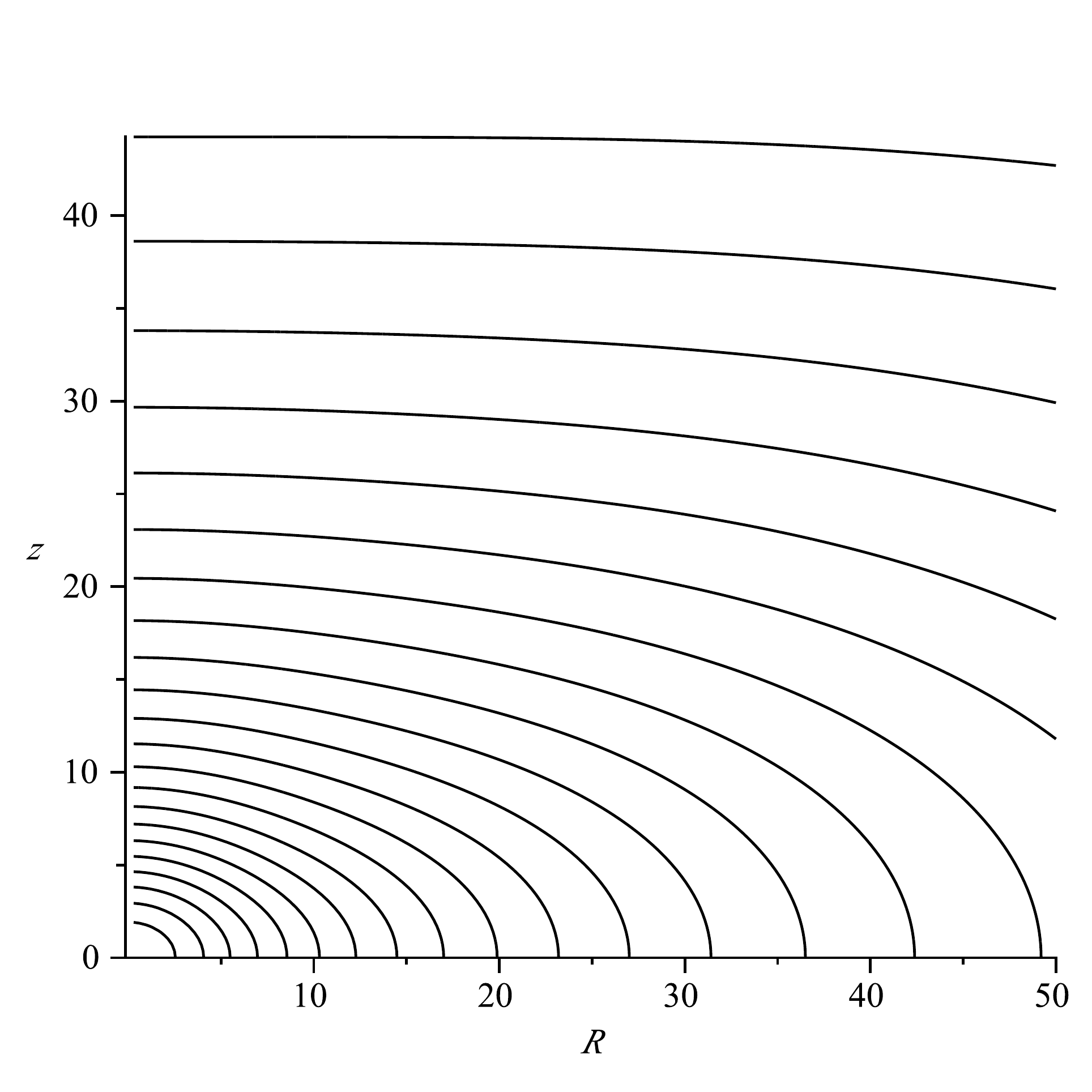}
}
\caption{Isolevels  for the  density distribution  resulting  from the
potential given by Eq.\,\ref{eq:abq}  with $(a,b,q)$=(5, 2, 1.3), (5,
2, 0.7), (0, 5, 0.5) from left to right.}
\label{f:pot-genera}
\end{figure*}

The Mestel  disk potential can be  generalized by  introducing a  core radius
$a$.  It gives (Evans  \& Collett  \cite{eva93}, and  quoted references
Rybicki, 1974, and Zang \cite{zan76})
$$ \phi(R,z)= \ln \left( \sqrt{R^2+(a+|z|)^2}+\left|\,a+|z|\,\right| \right),$$
with the resulting surface density
$$\Sigma(R)=\frac{1}{2\pi G\sqrt{R^2+a^2}},$$
and the rotation curve $v_c$, flat at large $R$, is given by
$$v_c(R)^2=\frac{R^2}{R^2+a^2+a\sqrt{R^2+a^2}}.$$

We generalize these Mestel disks  with a core by applying the Miyamoto
and Nagai  (\cite{MN}) method to  generalize the Plummer  spheroid and
Kuzmin disk (see also Baes \cite{bae09}). Thus we define the potential
\begin{equation}
\phi(R,z)= \ln \left(   
\sqrt{R^2+\left(a+\sqrt{b^2+z^2}\right)^2}
+~\left|\,a+\sqrt{b^2+z^2}\,\right|  
\,\,\right).
\label{eq:ab}
\end{equation}
It can be verified that this potential generates a positive spheroidal
density distribution.  With $b$=0 it may generate both a flat disk and
a spheroid. The  sum $(a+b)$ constrains the size of  the core, and $b$
the thickness of the disky part of the spheroid.

Potentials  given by Eqs.~\ref{eq:q} and  \ref{eq:ab} are peculiar
 cases ($q=0$ or $a=b=0$) of the more general potential

\begin{equation}\label{eq:abq}
\begin{array}{l}
\phi(R,z)  =   \\
 \ln    \left(    \sqrt{R^2+\left(a+  \sqrt{b^2+z^2}\right)^2   }     
    +\sqrt{q^2 R^2+\left(a+\sqrt{b^2+z^2}\right)^2}   \right)  .
\end{array}
\end{equation}

\noindent and the corresponding density is (if $b\ne 0$)

\begin{equation}\label{eq:densabq}
 \rho(R,z)= \frac{ q^2  m  + b^2 n}
{ 
\left(R^2+\alpha^2 \right)^{3/2}
 \left(q^2 R^2+\alpha^2 \right)^{3/2}
}
\end{equation}

\noindent with
\begin{equation}\label{}
 \begin{array}{l}
 m={\alpha}^{2} \left( 2\,{R}^{2}+{\alpha}^{2} \right) 
+ {\alpha\,{
b}^{2}{R}^{2} \left( {R}^{2}+{\alpha}^{2} \right) }
\, \beta^{-3}
+ z^2  R^4\, \beta^{-2}
\\
n= \left( {R}^{2}+\alpha^2  +\beta^{2}+\beta\,a \right)  \alpha^3
\,\beta^{-3}
\\
\alpha= a+\sqrt{b^2+z^2} 
\\
\beta=\sqrt{b^2+z^2}\,.

 \end{array}
 \end{equation}
 
If $a=0$, it may be shortly written as

\begin{equation}
 \rho(R,z)= 
 \frac{1}{4\pi G}
 \frac{  q^2  (R^2 + b^2 + z^2)^2  +
 \left( (q^2+1) R^2+ 2b^2+  2z^2    \right)b^2 }
{\left(R^2+b^2+z^2 \right)^{3/2}
 \left(q^2 R^2+ b^2+z^2 \right)^{3/2} } .
\end{equation}
When $b=0$ and $q\ne 0$,  it is the combination of a spheroid and a  flat disk
\begin{equation}
\rho(R,z)=\frac{1}{4\pi G}
 \frac{ q^2 \left( R^2+(a+|z|)^2 \right)^{1/2} }
{ \left(q^2 R^2+(a+|z|)^2 \right)^{3/2}},
\end{equation}
\begin{equation}
\Sigma(R)=
\frac{1}{2\pi G}
\frac{a}
{(R^2+a^2)^{1/2} (q^2 R^2+a^2)^{1/2} }.
\end{equation}

\noindent With  $q=1$, this is  the potential-density pair  defined by
Brada \& Milgrom (\cite{bra95}).

More  generally, we  obtain a  spheroid potential  producing  a rising
 velocity curve, flat  at large radii, the core  radius depending only
 on $q$  and on $a+b$.  The  parameter $q$ modifies  the flattening at
 large $R$  and $b$  the thickness  of the disky  part of  the density
 distribution.

Thus, this generalization introduces a new family of potential-density
pairs  with  general properties  similar  to  the  Miyamoto and  Nagai
potentials,  but with flat  circular velocity  curves at  large radii.
Figures  \ref{f:pot-genera} illustrate  isodensity contours  for three
such  potentials.   To  conclude,  we recall  another  three-parameter
family of potential-density pairs obtained by Zhao (\cite{zha96}).

\section{Anaytical estimate of the vertical velocity ellipsoid tilt}
\label{s:tilt}

We  examine  the  reliability  of  the Cuddeford  \&  Amendt  formulae
(Eqs.\,90-91 and E10, \cite{cud91})  and Amendt \& Cuddeford (Eq. 104,
\cite{amd91}) that predict the vertical tilt of the velocity ellipsoid
within an axisymmetric potential,  and we propose another formula that
is  more accurate  at greater  distances  from the  plane of  symmetry
$z=0$.   For various  potentials, we  compare the  expected  tilt from
these  two expressions  with the  tilt obtained  from  numerical orbit
integrations.

Cuddeford \& Amendt (\cite{cud91})  analyze the consecutive moments of
 the Boltzmann equation up to the 4th order by expanding these moments
 in Taylor  series assuming that $\sigma_v/{\rm  v}_{\bf circular}$ is
 small.  Combining these moment equations, they derive expressions for
 the  velocity moments and  obtain an  approximate expression  for the
 tilt    angle,   $\tan    2   \delta    =   \frac{2\,\sigma_{R,z}^2}{
 \sigma_{R,R}^2-\sigma_{z,z}^2   }$,  which   only   depends  on   the
 gravitational potential. They show  that their expression is exact in
 the  case of axisymmetric  St\"ackel potentials,  peculiar potentials
 for which  the tilt  only depends  on the potentials  but not  on the
 exact distribution function.   According to their initial hypothesis,
 the formula must be  generally valid when the considered distribution
 of stellar orbits covers  a sufficiently small domain (expecting that
 the covered  domain can be approximated with  a St\"ackel potential).
 This should be the case of nearly circular orbits with small vertical
 or radial oscillations.

 The  simplicity  of  the  result  obtained  by  Cuddeford  \&  Amendt
 (\cite{cud91}) makes it  extremely attractive. For instance Vallenari
 et al.  (2006) use it to  improve their recent galactic  model of star
 counts and kinematics, which  is dynamically self-consistent locally.

Inspired by their work, we  propose a more direct expression derivated
from  the  generic equation  (Ollongren  \cite{oll62}, Eq.\,7.2)  that
defines axisymmetric St\"ackel potentials:
\begin{equation}
\pm  z_0^2 = -\left(R^2-z^2\right) +\frac{R  z \left(  \phi_{R,R} -\phi_{z,z}
\right) + 3 \left( z \phi_R-R\phi_z\right) }{\phi_{R,z}} ,
\label{z0}
\end{equation}
with  a  positive  sign  for  prolate spheroidal  coordinates,  and  a
negative  one for  oblate  ones  (see de  Zeeuw,  \cite{zee85}, for  a
detailed description).

For a given potential $\phi$, we substitute $z_0$ from Eq.~\ref{z0} in
the expression for the velocity  ellipsoid tilt $\delta$ given by Hori
\&  Liu  (\cite{hor63}), which  is  exact  in  the case  of  St\"ackel
potentials:

\begin{equation}
\tan 2 \delta = \frac{2\,R\,z}{R^2-z^2\pm z_0^2} .
\label{tan}
\end{equation}

Combining  Eqs.~\ref{z0}  and  \ref{tan}  gives  an  estimate  of  the
velocity ellipsoid inclination within axisymmetric potentials. We note
that the  Amendt \& Cuddeford  (\cite{amd91}) formula could  have been
obtained  in  the same  way,  just  by  differentiating the  Ollongren
equation with respect to $z$ and replacing it within Eq.~\ref{tan}.\\

A  noticeable difference  between their  expression and  ours  is that
 their  formula is defined  at $z=0$  and is  exact for  the St\"ackel
 potential  only at  $z=0$, while  our expression  remains  exact (for
 St\"ackel potentials) at  any $z$. This may be why  our formula is in
 better  agreement  with the  results  of  the numerical  explorations
 described below.

Now, to  verify the reliability  of the estimated tilt  angle $\delta$
given  by Eqs.~\ref{z0}-\ref{tan},  and  the one  given  by Amendt  \&
Cuddeford  (\cite{amd91}) formulae,  we measure  the  tilt numerically
within a series of potentials.   For this purpose, we build stationary
distribution  functions with a  library of  2\,$10^7$ orbits  for each
model  using a  7th-order Runge-Kutta  (Fehlberg \cite{feh68})  with a
relative accuracy  $\epsilon=10^{-16}$.  Each orbit  is integrated for
80  rotations,   the  initial  conditions  being  drawn   from  a  Shu
distribution  function  (we  also  tried a  Dehnen  disk  distribution
function)  with $\sigma_R/{\rm  v}_{cir.}  =  0.4$  and $\sigma_z/{\rm
v}_{cir.}$ = 0.2 (we also consider the 0.2-0.1 and 0.6-0.3 pairs). One
point is selected per orbit in the last 40 rotations.

\begin{figure}[!htbp]
\resizebox{\hsize}{!}{
\includegraphics[angle=270]{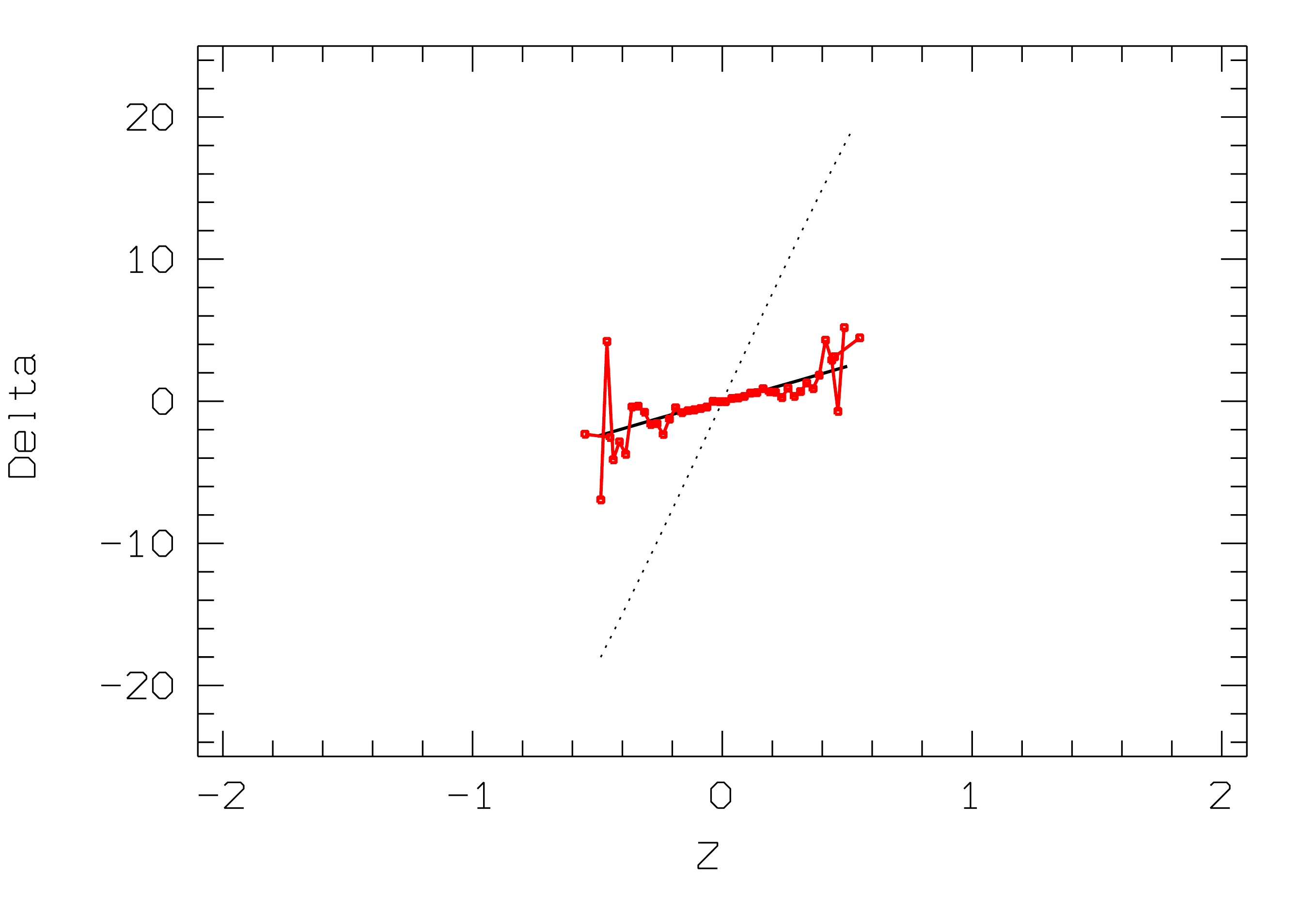}
\includegraphics[angle=270]{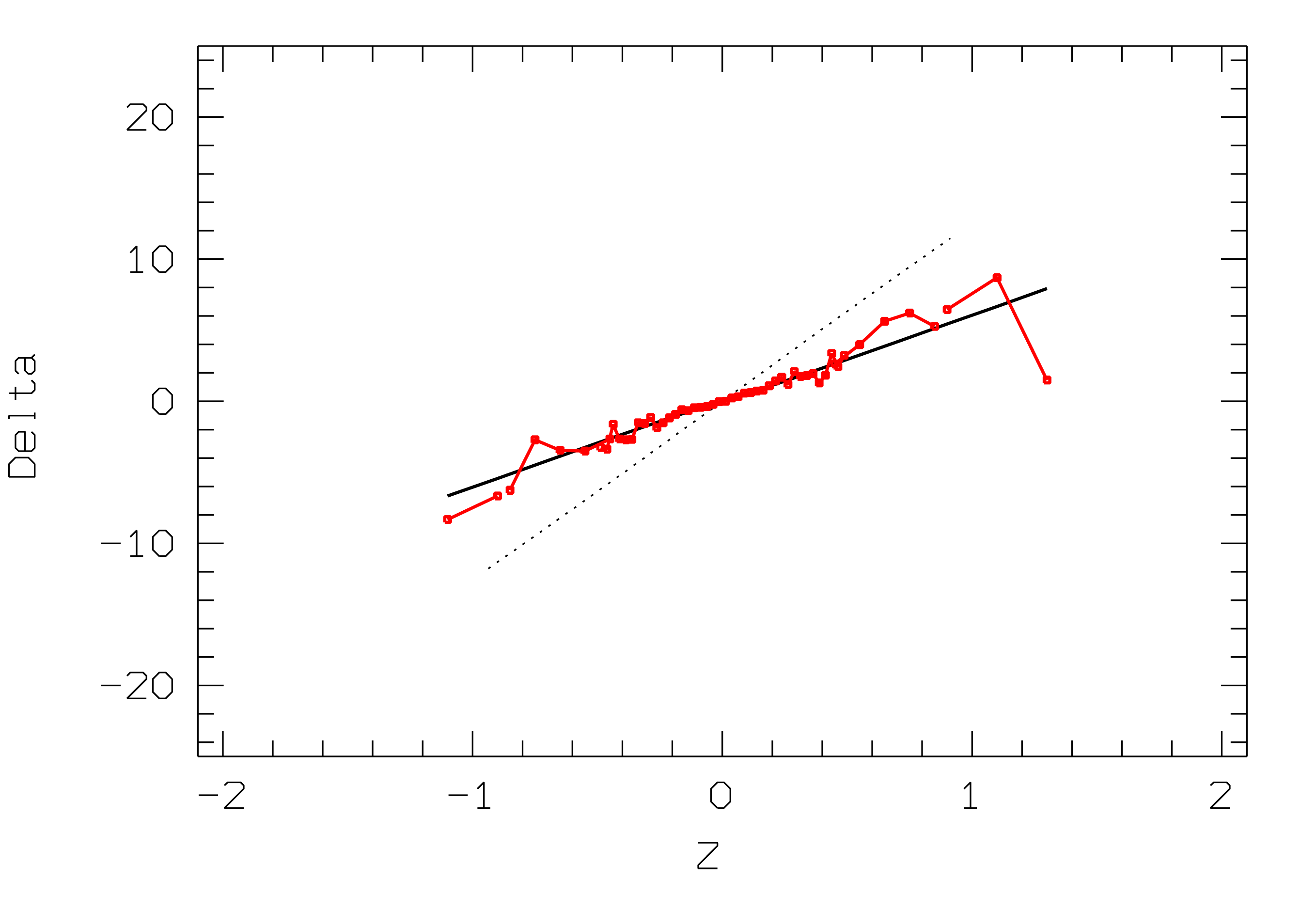}
}
\resizebox{\hsize}{!}{
\includegraphics[angle=270]{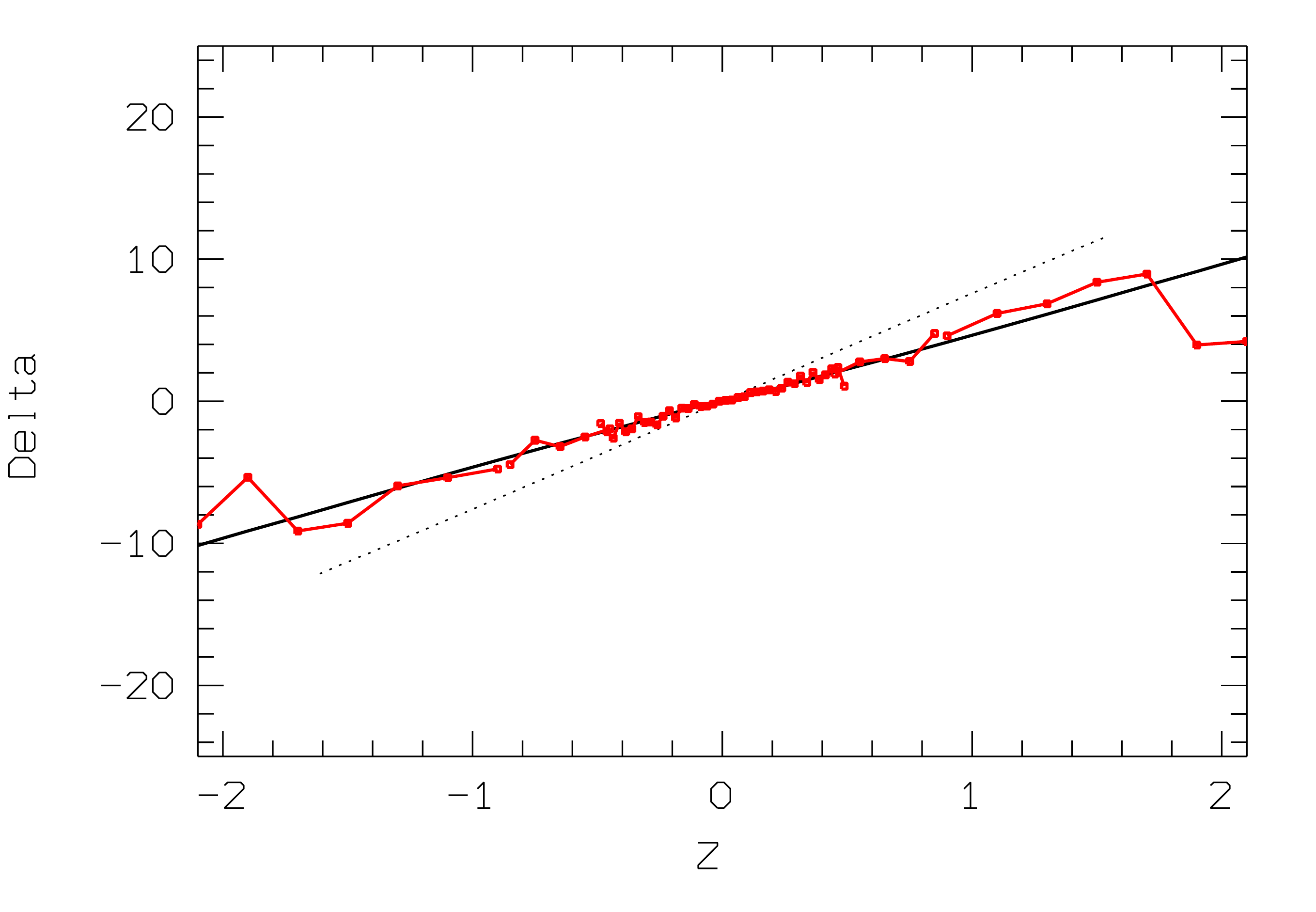}
\includegraphics[angle=270]{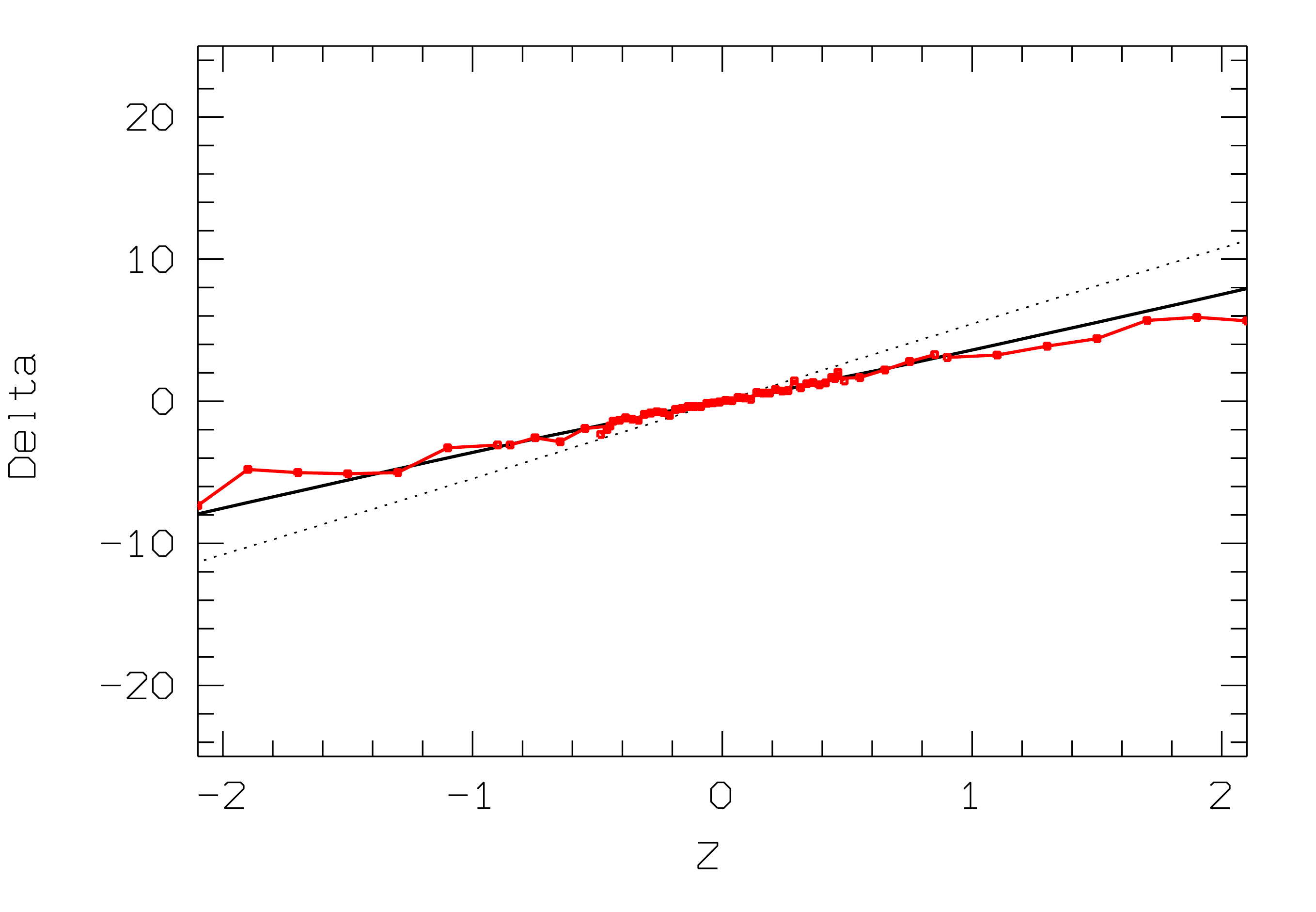}
}
\caption{Velocity ellipsoid  tilt angles in degrees  versus $z$ within
the  Brada  \&  Milgrom  (\cite{bra95}) Disk+Halo  potential  at  four
galactic  radii $R$=1.5,  4.5,7.5 and  10.5 (from  top left  to bottom
right). Red irregular  line: numerical determination.  Dark continuous
line:  analytical  estimate.  Dotted  line: tilt  within  a  spherical
potential.}
\label{D+H3}
\end{figure}

Figures \ref{D+H3}  show the inclination $\delta$  versus the distance
 $z$ from the  plane of symmetry in the case of  the Brada and Milgrom
 (\cite{bra95})  disk+halo potential  with a  core radius  $a=3$.  The
 tilt $\delta$ is plotted versus  $z$ at four different galactic radii
 $R$=1.5,  4.5,  7.5, and  10.5.   The  dark  line is  the  analytical
 estimate  from Eqs.  \ref{z0}-\ref{tan}, while  the red  line  is the
 result  of  numerical  orbit  integrations.   The  agreement  between
 numerical measures  and the analytical estimate is  satisfying in the
 range of  considered positions $R$  and $z$.  The  estimated vertical
 tilt  is also  shown in  Figure \ref{f:ivzq}:  at position  $R=8$ and
 $z=1$,  $\delta$  is  plotted   versus  the  flattening  $q$  of  the
 Eq.~\ref{eq:q}  potential.   The  agreement  is  satisfying  for  $q$
 between 0 and 1.

We explored other potentials: a  Kuzmin disk (a St\"ackel potential to
test our  numerical process), the  Mestel disk, the Brada  and Milgrom
potential,    a     logarithmic    flattened    potential,    combined
Kuzmin+logarithmic   halo  potentials.    For  these   potentials  our
analytical estimate give excellent results in agreement with numerical
determinations at a precision better than 1 degree at $z/R=0.15$.

With  the  slightly flattened  spherical  logarithmic potentials,  the
Amendt  \&  Cuddeford  (\cite{amd91})  formula is  also  an  excellent
estimate  for the  tilt (1  percent  difference at  $z/R=0.15$ with  a
potential axis ratio $q$=0.9).  However, with the Mestel potential, it
fails at any $z$ (a null  tilt is predicted), while it works correctly
with the Brada  and Milgrom potential, but only at  radius close to or
smaller than the core radius. \\

In conclusion,  our analytical  estimate of the  vertical tilt  of the
stellar  velocity ellipsoid  is  accurate for  various potentials  and
vertical distances  less than  $0.15R$.  It also  shows that at  1 kpc
above the  Galactic plane  at the solar  radius, $\delta$ varies  by 5
degrees depending  on whether the  potential is spherical or  the dark
matter component is flat.

However, for  an exponential  disk, this estimate  fails in  the range
$R<2$ or $>4$, in units of the scale length.  The situation is similar
for the  potential given by Eq.~\ref{eq:q} for  ``flattening" of $q=2$
and neighboring values.
An  explanation for  this deficiency  is that  the denominator  of the
r.h.s. of  Eq.~\ref{tan} is zero.   In that case, the  resulting value
for $z_0$ varies strongly with  $R$ or $z$.  Since Eq.~\ref{z0} is not
a St\"ackel fit, its domain of validity is determined by the condition
that the estimate for $z_0$  from Eq.~\ref{tan} does not vary strongly
over the  considered $(R,z)$-domain.  For deviations  from this limit,
more  sophisticated  methods  of  fitting  potentials  with  St\"ackel
potentials   are  required  (see   for  instance   de  Bruyne   et  al
\cite{bru00}) to  obtain more reliable estimates of  the vertical tilt
angle.

%

\end{document}